\documentclass[amssymb,amsmath,showpacs,superscriptaddress,preprint]{revtex4-1}
\usepackage{graphicx}
\usepackage{amsmath}
\usepackage{xcolor}
\usepackage{blindtext}
\usepackage[allcolors=blue]{hyperref}%
\newcommand{\BSS}[1]{{\color{black}#1}} 
\begin{document}
\title{Influence of oscillatory shear on nucleation in metallic glasses: a molecular dynamics study}
\author{Baoshuang Shang}
\email{shangbaoshuang@sslab.org.cn}
\affiliation{Songshan Lake Materials Laboratory, Dongguan 523808, China}
\author{No\"el Jakse}
\affiliation{Univ. Grenoble Alpes, CNRS, Grenoble INP, SIMaP, 38000 Grenoble, France}
\author{Pengfei Guan}
\affiliation{Beijing Computational Science Research Center, Beijing 100094, China}
\author{Weihua Wang}
\affiliation{Songshan Lake Materials Laboratory, Dongguan 523808, China}
\affiliation{Institute of Physics, Chinese Academy of Sciences, Beijing 100190, China}
\author{Jean-louis Barrat}
\email{jean-louis.barrat@univ-grenoble-alpes.fr}
\affiliation{Univ. Grenoble Alpes, CNRS, LIPhy, 38000 Grenoble, France}
\date{\today}
\begin{abstract}

  The process of crystal nucleation can be accelerated or retarded by ultrasonic vibration, which is particularly attractive for the addictive manufacture and thermoplastic forming of metallic glasses, however, the effect and mechanism of oscillatory loading on the nucleation process are still elusive. Here,
  by using molecular dynamics simulation, the changes in the time-temperature-transformation  (TTT) curve  under oscillatory external loading are systematically investigated in two typical binary alloys.
  A glass forming ability dependent response to the external loading  is found, and the shortest incubation time is insensitive to the external loading, while the corresponding temperature can be significantly shifted.
  Within the framework of classical nucleation theory, a fitting formula is proposed to describe the simulation data quantitatively. In contrast to stationary shear, the elastic stress, rather than the strain rate, is the key parameter to control the evolution of TTT curve under oscillatory loading.
  Furthermore, the model shows that oscillatory loading can decouple the mobility and nucleation in the deeply supercooled liquid, hence the deformation ability can be enhanced while the nucleation is suppressed, which is particularly helpful for the forming and manufacturing of metallic glasses.

\end{abstract}
\maketitle
\section{Introduction}

Understanding solidification phenomena during which a liquid morphs either into an amorphous or crystalline state is of fundamental importance as well as practical interest for materials manufacturing in industrial applications. It would enable a better control by defining the nonequilibrium pathway that determines the microstructure, both for natural and artificial materials\cite{Sosso2016crystal,kelton2010nucleation}. For amorphous materials,  avoiding nucleation during the quench is the key for improving the glass-forming ability (GFA) of the system\cite{greer2015new}. GFA is very high for many polymers or for silicates, due to the complex frustration between intrachain and interchain interactions, or to strong covalent bonds \cite{Jakse2012}. 

However, in metallic alloys, the metallic bond is less directional and weaker leading to a close-packing tendency, which makes the nucleation easier to happen during the quench. Since the seminal work of Turnbull \cite{Turnbull1949,turnbull1969under} the link between crystal nucleation and GFA still remains an open question that triggered many theoretical and experimental works \cite{Turnbull1949,turnbull1969under,Inoue2000stabilization,PhysRevLett.91.115505,tang2013anomalously,zhang2013computational,Russo2018,Nie2020connecting,Hu2020,PhysRevX.9.031016}. Such a relationship may find its roots at the atomic scale  in terms of the variety of favoured local structures \cite{Ronceray2011}, competing short-range orders \cite{Jakse2003,Becker2020,Pedersen2021}, or interplay between chemical and fivefold symmetry orderings \cite{Jakse2008,Pasturel2017,Tang2018} . The overall objective is to find a connection between nucleation or growth parameters such as liquidus temperature, atomic size or chemical potential with the GFA of metallic glasses (MGs). In contrast, there are fewer works that focus on the effect of external loading on the GFA of metallic glasses\cite{PhysRevE.91.020301}. 

Recently, new fabrication processes such as rheological processing\cite{Johnson2011}, thermoplastic\cite{ADMA200902776}, and additive manufacturing\cite{Sohrabi2021} were designed to improve the GFA of MGs by reheating and forming MGs in the supercooled liquid or glassy state. This immediately raises the question of the evolution of GFA with the external loading. 
As studied both in the simulation and experimental works\cite{PhysRevE.87.062307,Richard2015,PhysRevE.82.021505,PhysRevE.93.042803}, the nucleation process can be retarded or accelerated in a sheared liquid, that is to say, the GFA could be changed in stationary shear flow. 
\BSS{Beside the steady  shear deformation,  oscillatory shear as a specific deformation mode is widely used in the study of metallic glasses and supercooled liquids, such as dynamic relaxation or mechanical annealing\cite{Priezjev2018,PhysRevB.90.144201,Wang2020revealing,PhysRevLett.122.105501}, especially
the oscillatory deformation induced by ultrasonic vibration can be applied in the forming of MGs\cite{Li2014,Ma2019}} and it can also accelerate the crystallization of MGs\cite{PhysRevLett.95.245501}. In general,  oscillatory shear allows a glass to access a broader range of energies \cite{Priezjev2021}, but how the parameters of oscillatory deformation affects the nucleation process, and how the TTT curve changes under oscillatory deformation,  are still open questions.

In the present paper, by using molecular dynamics simulation (MD), we investigate the homogeneous nucleation behavior of  alloys under oscillatory shear conditions, with amplitude and period as varying parameters.  For this purpose, two typical binary alloys, namely Ni$_{50}$Al$_{50}$ and Cu$_{23}$Zr$_{77}$, were considered as they both have a stable underlying crystalline structure, but display a significantly different GFA \cite{Jakse2015,tang2013anomalously}. Interestingly, their crystallization times are reachable by brute force MD, giving access to a detailed description of the structure and dynamics. We find that the oscillation can significantly affect the time-temperature-transformation (TTT) curve for both alloys. 
More precisely, the shortest incubation time of the TTT curves is insensitive to the various external oscillatory loading conditions, while the corresponding temperature does depend on them. 
We further show that the nucleation process can be controlled by two loading parameters, which differ significantly from those obtained in stationary shear flow.  Finally, we rationalize our results by proposing a simple phenomenological formula that describes these observations, in which the elastic stress associated with the external loading is the main parameter.

\section{Methods}
\subsection{\BSS{Initial sample preparation}}

Molecular Dynamics simulations were performed by using the open-source software \textsc{LAMMPS} \cite{Thompson2022}. A number of atoms $N=8192$ for Ni$_{50}$Al$_{50}$ and $N = 10000$ for Cu$_{23}$Zr$_{77}$ were placed randomly in a cubic simulation box subject to the standard periodic boundary conditions (PBC) in the three directions of space. Interatomic interactions were taken into account through the semi-empirical potentials based on the embedded atom model (EAM) for Ni$_{50}$Al$_{50} $ \cite{PurjaPun2009development} as well as Cu$_{23}$Zr$_{77}$ \cite{Mendelev2009development}. Integration of the equations of motion were solved using Verlet's algorithm in the velocity form with a time step of $2$ fs. The thermodynamic conditions were controlled by means of the isothermal–isobaric (NPT) ensemble at ambient pressure using the Nosé-Hoover thermostat and barostat \cite{nose1984unified,martyna1994constant}.

The samples were first equilibrated at $3000$ K for $1$ ns, where the temperature was far above the liquidus temperature ($T_L$ ) of each system.
The investigated samples were prepared by fast quenching from $3000$ K to a target temperature with a fast cooling rate around $10^{13}$ K/s at constant pressure. During cooling, a configuration is taken and saved at each target temperature to further construct the TTT curves for various loading conditions along isotherms. 

\subsection{\BSS{Oscillatory shear deformation}}

Subsequently, during the isothermal homogeneous nucleation, the simulation box was deformed by using a sinusoidal shear deformation along the $x$ direction. 
The time dependent  oscillatory shear  strain is $\gamma(t)=\gamma_A \sin (2\pi {t}/{t_p})$, where $\gamma_A$ is the amplitude and $t_p$ is the oscillation period. 
During the deformation, we solved the SLLOD equations\cite{Edwards2006validation} of motion with the Nos\'{e}-Hoover thermostat and PBC in all dimensions, keeping the volume and temperature constant (NVT ensemble). 

\subsection{\BSS{Characterization of crystallization}}

We monitored the degree of crystallization at the target temperature by calculating the parameter $F_6$ \cite{goodrich2014solids}.
The crystal bond of two neighbouring atoms i and j was constructed by the scalar product
\begin{equation}
  S_6(i,j) \equiv \frac{\sum_{m=-6}^{6} q_{6m}(i) \cdot q_{6m}^{*}(j)}{\sqrt{\sum_{m=-6}^{6}q_{6m}(i)\cdot q_{6m}^{*}(i)}\sqrt{\sum_{m=-6}^{6}q_{6m}(j) \cdot q_{6m}^{*}(j)}}
\end{equation}
where $q_{6m}$ is the standard bond-orientations order parameter\cite{PhysRevB.28.784} and $q_{6m}^{*}$ is the corresponding complex conjugate. 
We defined the bond between neighbour atoms $i$ and $j$ as a crystalline bond when $S_6(i,j)>0.7$.
Summing over all the crystal bonds of atom $i$, the local degree of crystallinity for atom $i$ is written as\cite{Auer2004numerical,Russo2012microscopic}
\begin{equation}
  f_6 =\frac{1}{N_{c}(i)}\sum_{j \in \{N_{c}(i)\}} \Theta (S_6(i,j)-0.7)
\end{equation}
Where $\Theta(x)$ is the step function and $N_{c}(i)$ is the number of neighbours  of atom $i$. 
The parameter $F_6$ is obtained by averaging $f_6$ { over all atoms of the simulation box}. 
For a perfect crystal, $F_6=1$ and in a liquid state, $F_6$ is small and approaches zero.
For all data points reported, five independent simulations were used to improve the statistics and estimate the error bars.
\subsection{\BSS{Dynamical modulus analysis}}

\BSS{Dynamical modulus analysis is widely used in the deformation and relaxation mechanism of amorphous solid and supercooled liquids\cite{PhysRevB.90.144201,Wang2020revealing,PhysRevLett.122.105501}.
Here, we used an oscillatory shear $\gamma=\gamma_A \sin(2\pi t/t_p)$ to obtain the dynamic modulus spectrum at a given temperature. The storage modulus $G'(t_p)$ and loss modulus $G''(t_p)$ can be obtained from the measured shear stress as
\begin{eqnarray}
  G'(t_p)=\frac{2}{N t_p \gamma_A} \int_{0}^{N t_p} \sin(2\pi t/t_p)\sigma(t) dt \\
  G''(t_p)=\frac{2}{N t_p \gamma_A} \int_{0}^{N t_p} \cos(2\pi t/t_p)\sigma(t) dt
  \label{eqn:losstore}
\end{eqnarray}
Here  $\sigma(t)$ is the shear stress along the shear strain direction, and $N$ is the number of cycles, here  $N=100$. We obtained the dynamical modulus spectrum of Cu$_{23}$Zr$_{77}$ at 700 K with $\gamma_A=0.06$, which will be used in the following.
}

\section{Results and Discussion}
\subsection{\BSS{Time-temperature-transformation (TTT) curve}}

\begin{figure}[!htpb]
  \centering
  \includegraphics[width=0.8\textwidth]{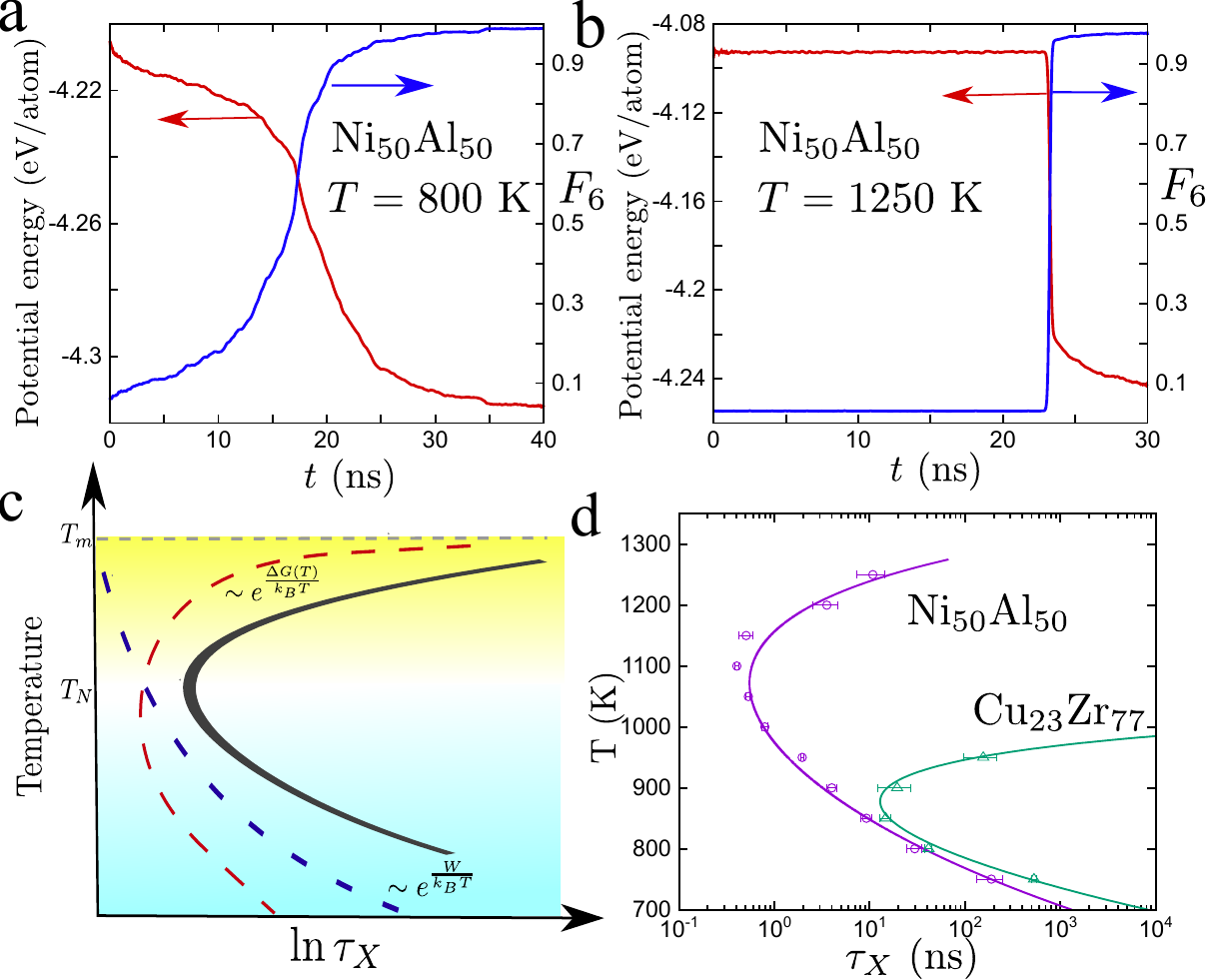}
  \caption{ 
  \textbf{Homogeneous nucleation from a supercooled liquid.} 
  \textbf{a.,b.} {Potential energy of supercooled  Ni$_{50}$Al$_{50}$ evolving with waiting time as the degree of crystallization ($F_6$) increases}.
  \textbf{a.} shows the typical nucleation behaviour in the deeply supercooled liquid regime (temperature below the nose temperature $T_{N}$ of the quiescent state) , \textbf{b.} Typical nucleation behaviour in the moderately supercooled regime (above $T_{N}$).    
  \textbf{c.} Schematic of time-temperature-transformation (TTT) curve. The temperature above $T_N$ (yellow region) is the moderately supercooled liquid regime where the thermodynamic term dominates the nucleation, and the temperature below $T_N$ (cyan region) is the deeply supercooled liquid regime, where the kinetic factor dominates the nucleation.
  The dashed lines represent the kinetic factor (${W}/{k_B T}$) and thermodynamic factor  (${\Delta G(T)}/{k_B T}$), respectively. The liquidus temperature $T_L$ and nose temperature $T_N$ are also shown.
  \textbf{d.}  TTT curve of Ni$_{50}$Al$_{50}$ and Cu$_{23}$Zr$_{77}$, the incubation time $\tau_X$ is defined as the threshold $F_6=0.5$.
  The data points are obtained from the MD simulations, and the uncertainties are estimated from five independent MD results, \BSS{the solid curves are fitted by the classical nucleation formula. }
  }
  \label{fig:1}
\end{figure}

According to classical nucleation theory (CNT), the nucleation process from a supercooled liquid can be ascribed to the competition between kinetic and thermodynamic factors\cite{Turnbull1949,turnbull1969under,cavagna2009supercool}, and the incubation time $\tau_X$ can be written as:
\begin{equation}
  \tau_X = A e^{\frac{W}{k_B T}} e^{\frac{\Delta G(T)}{k_B T}},
  \label{eqn:1}
\end{equation}
\BSS{
where $A$ is a prefactor,  ${W}/{k_B T}$ is the kinetic factor which describes the difficulty for atomic attachment to the crystalline nuclei. As temperature decreases, the kinetic factor increases.  $W$ is the kinetic barrier. Both $A$ and $W$ are supposed to be insensitive to temperature.
${\Delta G(T)}/{k_B T}$ is the thermodynamic factor,  which describes the difficulty for generating a critical crystalline nucleus through thermal fluctuations.} $\Delta G(T)$ is  is determined by the balance between the chemical potential difference and the liquid-crystal interfacial energy\cite{turnbull1969under}.
$\Delta G(T)$ varies with temperature, increasing rapidly near the liquidus temperature ($T_L$) and then decreasing with temperature. 
Approaching the liquidus temperature, the \BSS{thermodynamic factor} is larger than the \BSS{kinetic factor}, and the nucleation process is dominated by the thermodynamic barrier ($\Delta G(T)$). 
As temperature decreases, the influence of the \BSS{thermodynamic factor} declines but the \BSS{kinetic factor} becomes dominant. 
Below a certain temperature, the \BSS{kinetic factor} becomes a dominant factor in the nucleation process.
Therefore the nucleation time of the supercooled liquid is non-monotonically dependent on temperature, and with a transition from thermodynamically to kinetically controlled nucleation as temperature decreases.
This non-monotonic behaviour is usually illustrated by the time-temperature-transformation (TTT) curve (Figure \ref{fig:1}). 
At the ``nose temperature" ($T_N$), the incubation time ($\tau_N$) is minimum (Figure \ref{fig:1} (c)). It is worth noting that the value $\tau_N$ can be often taken as a quantitative measure of the GFA of the alloy\cite{Johnson2016quantifying}.

Figure \ref{fig:1} (a) shows a typical nucleation event in the deeply supercooled liquid regime (below the nose temperature $T_N$) for Ni$_{50}$Al$_{50}$. One sees that the potential energy decreases gradually as the degree of crystallization ($F_6$) increases. This contrasts with the 
slightly supercooled case (above $T_N$), shown in Figure \ref{fig:1} (b), where the crystalline nuclei are difficult to form. 
However, once an embryo attains the critical radius, the resulting nucleus grows rapidly thanks to the high  atomic mobility. 
As a result, both the potential energy and the degree of crystallization change abruptly during the nucleation event. 
Figure \ref{fig:1} (c) schematically illustrates the contribution of the kinetic and thermodynamic factors to the TTT curve.
The incubation time $\tau_X$ is defined as the time when the degree of crystallization $F_6$ reaches the threshold $0.5$. Other thresholds were also investigated, with  results  qualitatively consistent with those obtained for $F_6=0.5$. A more detailed discussion  of the TTT curve is given in the Appendix.
Figure \ref{fig:1} (d) shows incubation times versus temperatures in the quiescent state for the two alloys.
The nose temperature of the quiescent state \BSS{($T_N^Q$)} is about $1080$ K and $870$ K for Ni$_{50}$Al$_{50}$ or Cu$_{23}$Zr$_{77}$, respectively. 
For Cu$_{23}$Zr$_{77}$, the incubation time at the nose temperature of the quiescent state \BSS{($\tau_N^Q$)} is larger than for Ni$_{50}$Al$_{50}$ (more than one order of magnitude), indicating that the GFA of Cu$_{23}$Zr$_{77}$ is significantly higher than that of Ni$_{50}$Al$_{50}$ \cite{Johnson2016quantifying}.

\subsection{\BSS{Homogeneous nucleation under oscillatory shear}}

\begin{figure}[!htpb]
  \centering
  \includegraphics[width=0.8\textwidth]{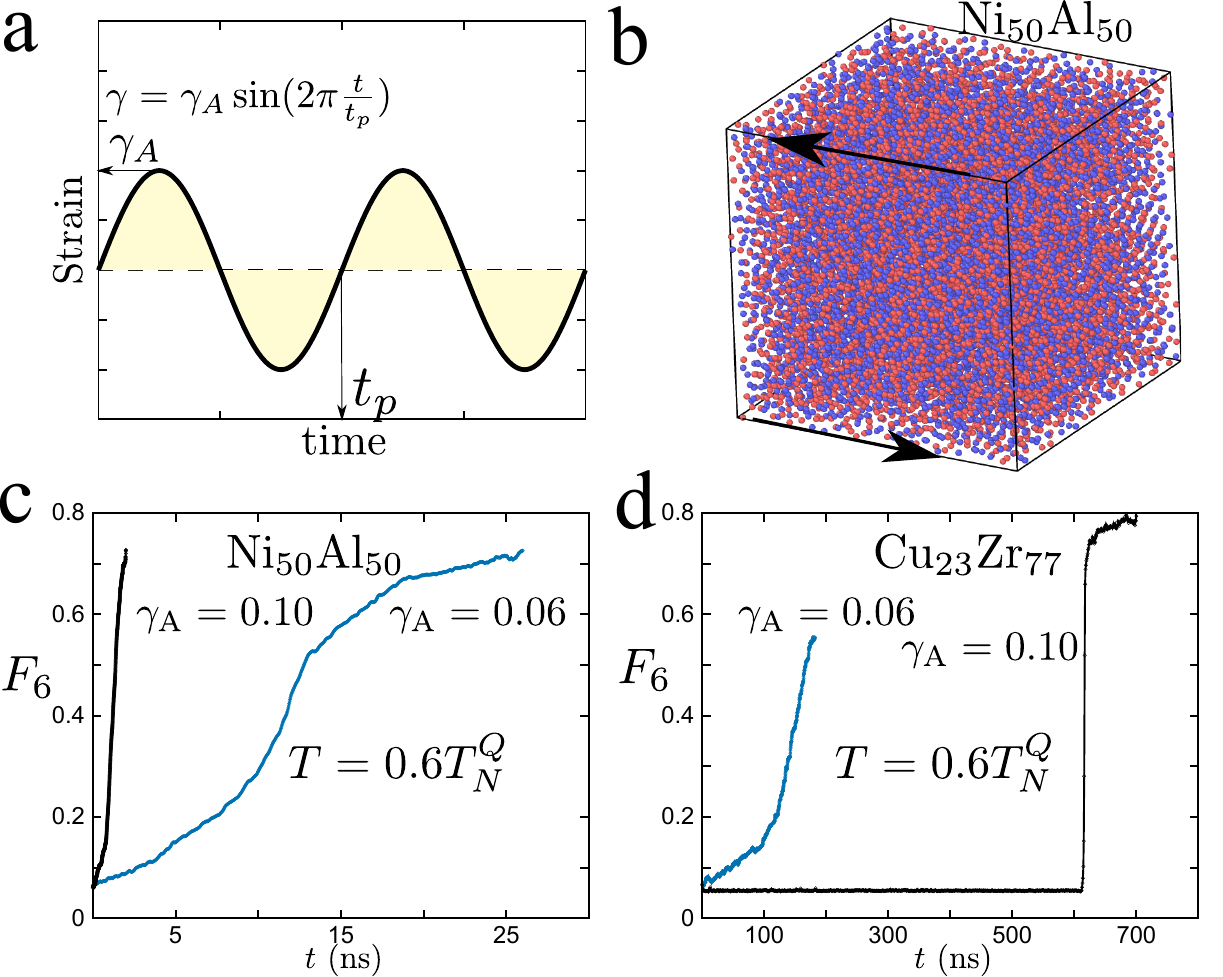}
  \caption{\textbf{Homogeneous nucleation under oscillatory shear.} 
  \textbf{a.} Time evolution of shear strain during oscillatory shear deformation,  the period $t_p$ and amplitude $\gamma_A$ are indicated. 
  \textbf{b.} Snapshot of the atomic configuration of Ni$_{50}$Al$_{50}$ under shear, the deformation direction is illustrated.
  \textbf{c.,d} Nucleation  of Ni$_{50}$Al$_{50}$ and Cu$_{23}$Zr$_{77}$  under oscillatory shear $\gamma_A=0.6$, or $\gamma_A=0.1$ and {$t_p=0.02$ ns} at a temperature $\sim 0.6 T_{N}^Q$, {with} $600$ K for Ni$_{50}$Al$_{50}$ and $500$ K for Cu$_{23}$Zr$_{77}$, respectively.
  }
  \label{fig:2}
\end{figure}

Next, we used a sinusoidal deformation  to investigate the effect of external loading on the TTT curve.
Figure \ref{fig:2} (a) illustrates of the  two control parameters in the loading: the  amplitude of the shear strain $\gamma_A$ and its period $t_p$.  
A snapshot of an atomic configuration showing the loading direction is displayed in Figure \ref{fig:2} (b).
At a given temperature, which we choose first to be in the deeply supercooled regime $(\sim 0.6 T_{N}^Q)$,
we impose the  period  $t_p=0.02$ ns and investigate first the effect of the loading amplitude on the nucleation process.
Figure \ref{fig:2} (c), (d) illustrates the nucleation process under oscillatory shear with $\gamma_A=0.06$ or $0.1$ and $t_p=0.02$ ns at the  temperature $\sim 0.6 T_{N}^Q$.
Interestingly, although the control parameter $t_p$ and relative temperatures are the same in both Ni$_{50}$Al$_{50}$ and Cu$_{23}$Zr$_{77}$, the nucleation process with different amplitude strains is still notably distinct.
For Ni$_{50}$Al$_{50}$, both for $\gamma_A=0.06$  and $\gamma_A=0.1$, the degree of crystallization increases gradually, and  shows a behavior typical of the nucleation process in a deeply supercooled liquid (below $T_N$).
The incubation time decreases when increasing $\gamma_A$ (Figure \ref{fig:2} (c)).
This suggests  that, as $\gamma_A$ increases, the kinetic factor decreases, but is still the dominant term in this region of the TTT curve in Ni$_{50}$Al$_{50}$.
For Cu$_{23}$Zr$_{77}$, a completely different behavior is observed. For $\gamma_A=0.06$, the nucleation process still retains the  features of the deeply supercooled liquid. For $\gamma_A=0.1$, however, the nucleation process is  retarded and has a behaviour  typical of  nucleation from a moderately  supercooled liquid (above $T_N$). 
This suggests a transition from kinetic control to thermodynamic control as the strain amplitude is increased at fixed temperature in Cu$_{23}$Zr$_{77}$.

\subsection{\BSS{The effect of oscillatory loading within the framework of CNT}}
\BSS{
The oscillatory shear can  change both the kinetic and thermodynamic factors. Following the work  of Turnbull \cite{turnbull1969under}, we incorporate the effect of external loading into the CNT formula, so that  the incubation time $\tau_X$ can be rewritten as
}
\BSS{
\begin{equation}
  \tau_X(T,\gamma_A,t_p) = \tau_0 e^{\delta(\gamma_A,t_p)} e^{\frac{W(\gamma_A,t_p)}{k_B T}} e^{\frac{\Gamma(\gamma_A,t_p)}{k_B T (1-T/T_L(\gamma_A,t_p))^2}}
  \label{eqn:cnt}
\end{equation}
Where by convention $\tau_0 \equiv 10^{-12}$ s. $k_B$ is Boltzmann constant and $T$ is the temperature. There are four parameters in  equation \ref{eqn:cnt}, which are independent of temperature but sensitive to external loading: the prefactor $\delta(\gamma_A,t_p)$ , the kinetic barrier $W(\gamma_A,t_p)$, the liquid-crystal interfacial parameter $\Gamma(\gamma_A,t_p)$ and the liquidus temperature $T_L(\gamma_A,t_p)$. 
Those parameters can be obtained by fitting equation \ref{eqn:cnt} to the simulation results.
}

\begin{figure}[!htpb]
  \centering
  \includegraphics[width=0.8\textwidth]{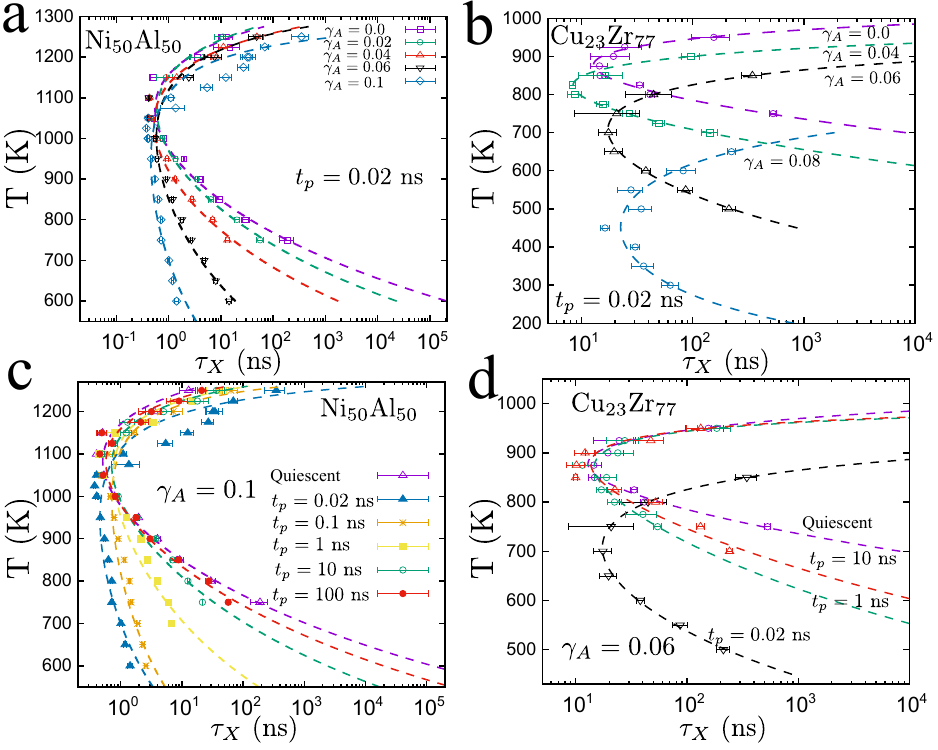}
  \caption{
  \textbf{The role of oscillatory loading on the TTT curves.} 
  \textbf{a,b} The TTT curves for various strain amplitudes and fixed the period {$t_p=0.02$ ns} for Ni$_{50}$Al$_{50}$ and Cu$_{23}$Zr$_{77}$, respectively. 
  \textbf{c,d} The TTT curves for various periods and fixed strain amplitude $\gamma_A=0.1$ and $\gamma_A=0.06$ for Ni$_{50}$Al$_{50}$ and Cu$_{23}$Zr$_{77}$, respectively.
  The external loading is an oscillatory simple shear $\gamma = \gamma_A \sin (2\pi t/t_p)$. 
  \BSS{The dashed lines are fitted by equation \ref{eqn:cnt} with the corresponding simulation data.}
  }
  \label{fig:3}
\end{figure}

\begin{figure}[!htbp]
  \centering
  \includegraphics[width=0.8\textwidth]{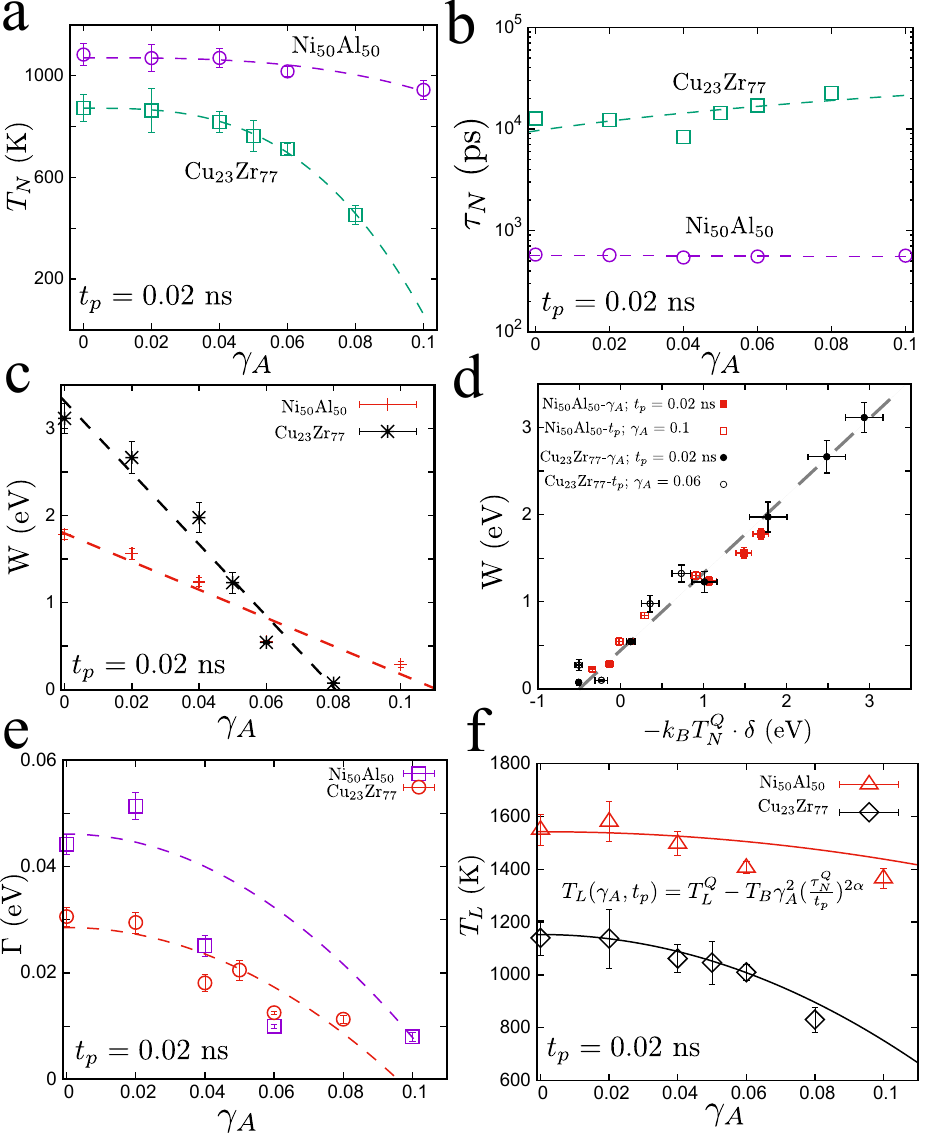}
  \caption{ \BSS{
  \textbf{Evolution of the fitting parameters with external loading.}
  \textbf{a.} The nose temperature $T_N$ versus the strain amplitude  $\gamma_A$. 
  \textbf{b.} The shortest incubation time $\tau_N$ of nose temperature versus amplitude strain $\gamma_A$ both for {Ni$_{50}$Al$_{50}$ and Cu$_{23}$Zr$_{77}$}. 
  \textbf{c.} The kinetic barrier $W$ changes with amplitude strain $\gamma_A$, both for {Ni$_{50}$Al$_{50}$ and Cu$_{23}$Zr$_{77}$} with {$t_p=0.02$ ns}. 
  \textbf{d.} The correlation between kinetic barrier $W$ and the prefactor $-k_B T_{N}^{Q} \cdot \delta$. 
  $T_{N}^{Q}$ is the nose temperature of quiescent state. 
  \textbf{e.} The thermodynamic barrier $\Gamma$ changes with $\gamma_A$, both for {Ni$_{50}$Al$_{50}$ and Cu$_{23}$Zr$_{77}$} with {$t_p=0.02$ ns}.
  \textbf{f.} The liquidus temperature $T_{L}$ changes with $\gamma_A$, both for Ni$_{50}$Al$_{50}$ and Cu$_{23}$Zr$_{77}$ with $t_p=0.02$ ns, the solid line is predicted by the effective temperature model, the equation is shown in the panel where $T_L^Q$ is the liquidus temperature at quiescent state, $T_B$ is a material parameter from effective temperature model.
  The dashed lines are guided for the eye. }
  \label{fig:4}
  }
\end{figure}

\BSS{
We systematically obtained the TTT curves for different strain amplitudes, keeping the period fixed at $t_p=0.02$ ns on the one hand, and for different periods, with a strain amplitude $\gamma_A=0.06$ or $\gamma_A=0.1$, on the other hand.
Figure \ref{fig:3} shows the corresponding results both in Ni$_{50}$Al$_{50}$  (panel (a),(c)) and Cu$_{23}$Zr$_{77}$ (panel (b),(d)).
The simulation data can be well fitted by equation \ref{eqn:cnt}, and the TTT curve is shown as a dashed line. 
The corresponding parameters are shown in figure \ref{fig:4}.
} 

For Ni$_{50}$Al$_{50}$, the nucleation process is quite fast, and the shortest incubation time $\tau_N$ is less than 1 ns.
Both the nose temperature $T_N$ and the corresponding incubation time $\tau_N$ are insensitive to the strain amplitude and period (\BSS{Figure \ref{fig:4} (a),(b)}).  
Furthermore, above $T_N^Q$, the nucleation process is slightly retarded, and below $T_N^Q$, the nucleation process is significantly accelerated (\BSS{Figure \ref{fig:3} (a),(c)}). 
The main effects of the external oscillatory loading, obtained either by increasing $\gamma_A$ or by decreasing $t_p$, are that (i) the kinetic factor is significantly decreased \BSS{(Figure \ref{fig:4})}, and (ii) the thermodynamic factor is slightly enhanced.

In contrast, for Cu$_{23}$Zr$_{77}$, the nucleation process is relatively slow, and $\tau_N$ is larger than 10 ns.
For a fast period $t_p=0.02$ ns, Figure \ref{fig:3} (b) \BSS{and Figure \ref{fig:4} (b)} shows that $\tau_N$ is almost insensitive to the external loading, but the nose temperature $T_N$ decreases when $\gamma_A$ increases \BSS{(figure \ref{fig:4} (a))}.
The shift of the  nose temperature of the TTT curve when increasing  $\gamma_A$ in Cu$_{23}$Zr$_{77}$ is consistent with the observation of the transition from kinetic control to thermodynamic control in Figure \ref{fig:2} (d).
This suggests that the thermodynamic factor is significantly increased by the external loading, and the balance between the kinetic and thermodynamic factors shifts to a lower temperature. 
In contrast, for a slow oscillation period ($t_p=1$ ns and $10$ ns in Figure \ref{fig:3} (d)), the effect of external loading is similar to the case of  Ni$_{50}$Al$_{50}$ system in Figure \ref{fig:3} (c). 

Hence the shift in the nose temperature behavior in Cu$_{23}$Zr$_{77}$ is dependent on the competition between the external period $t_p$ and the nose incubation time $\tau_N^Q$. For a slow period, when the ratio $\tau_N^Q/t_p$ is approaching  the situation of Ni$_{50}$Al$_{50}$, the nose temperature is insensitive to the amplitude of the strain. 
For faster oscillations, the ratio $\tau_N^Q/t_p$ is large, and the thermodynamic factor will be increased notably by the  strain, so that the nose temperature will be shifted to a lower temperature (\BSS{Figure \ref{fig:4} (a)}).

\subsection{\BSS{Shear melting and effective temperature model}}
\begin{figure}
  \centering
  \includegraphics[width=0.6\textwidth]{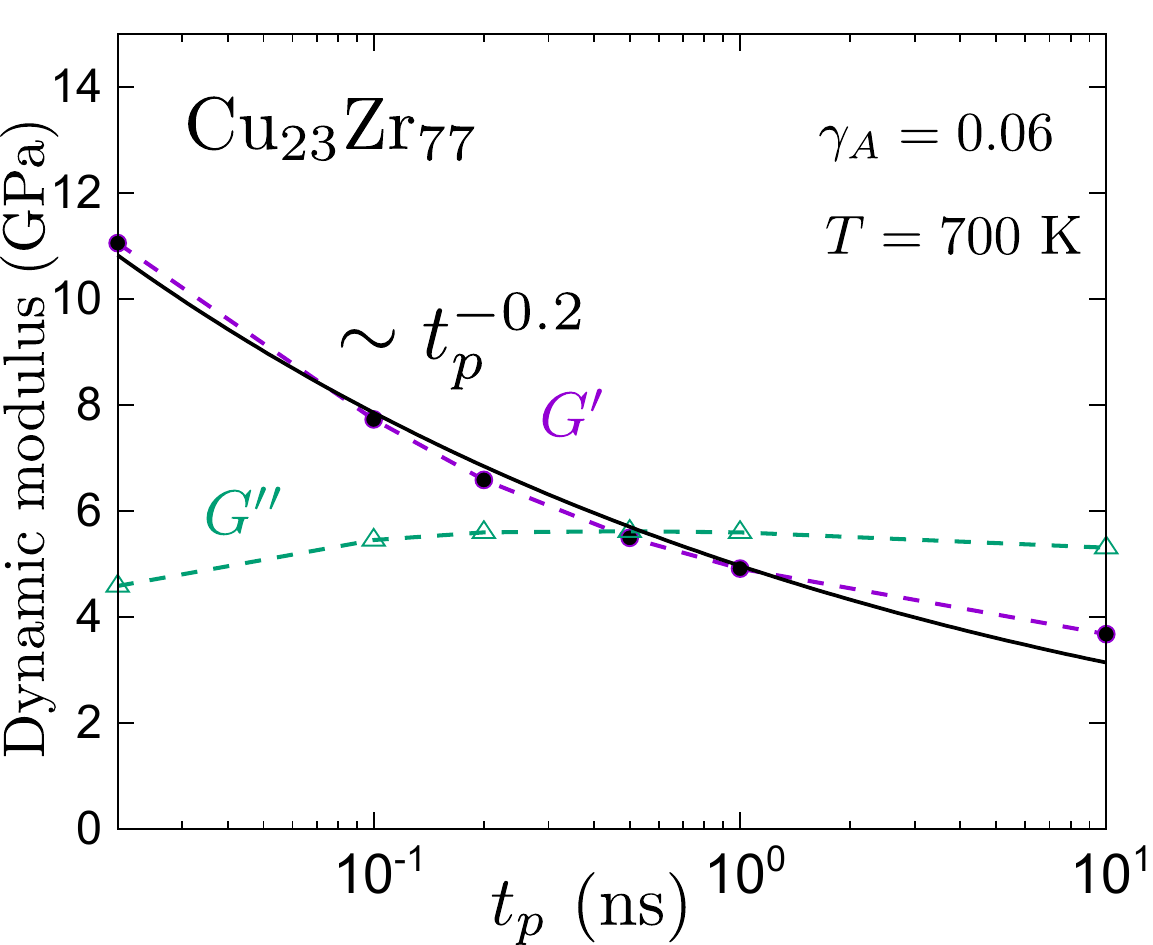}
  \caption{\BSS{Dynamic modulus spectrum of Cu$_{23}$Zr$_{77}$ with $\gamma_A=0.06$ at 700 K, $G'$ is storage modulus and $G''$ is loss modulus, the solid line illustrates $\sim t_p^{-0.2}$. The dashed lines are guided for the eye.}}
  \label{fig:5}
\end{figure}
\begin{figure}
  \centering
  \includegraphics[width=0.8\textwidth]{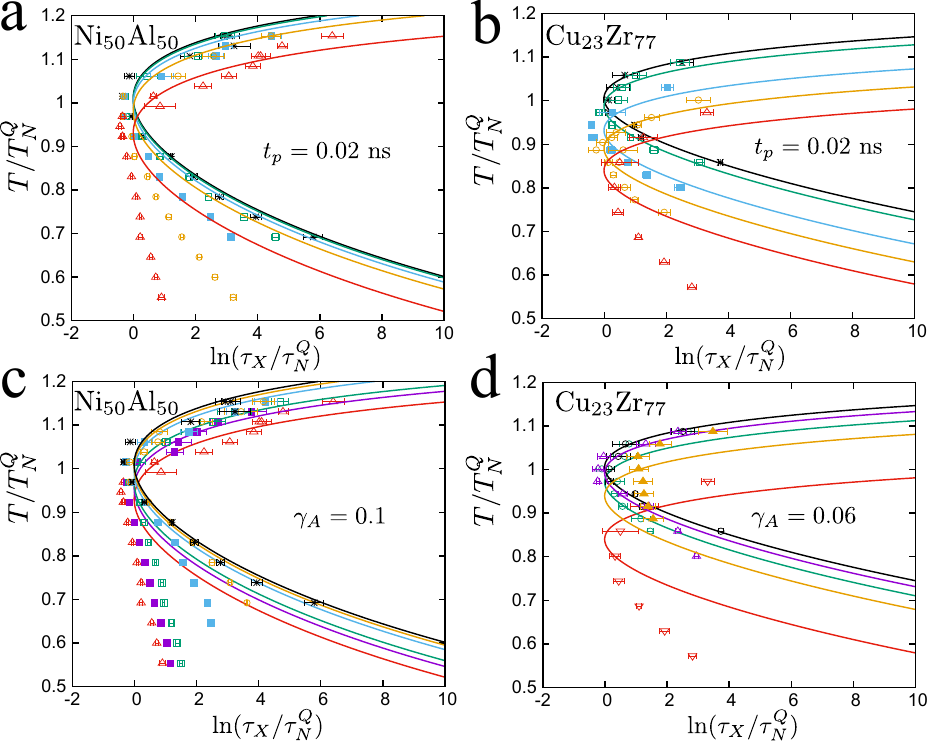}
  \caption{ 
  \BSS{ \textbf{Comparison between the effective temperature model and simulation results}
  \textbf{a,b} Evolution of TTT curves with strain amplitude  $\gamma_A$ for Ni$_{50}$Al$_{50}$ and Cu$_{23}$Zr$_{77}$, respectively, the period is $t_p=0.02$ ns.
  \textbf{c,d} Evolution of TTT curves with various periods for Ni$_{50}$Al$_{50}$ with $\gamma_A=0.1$, and for Cu$_{23}$Zr$_{77}$ with $\gamma_A=0.06$. 
  The points  represent the simulation results, and the solid curves are calculated from equation \ref{eqn:A3}.
  }
  }
  \label{fig:6}
\end{figure}

\BSS{
Generally, the liquidus temperature $T_L$ can be changed by an external loading, as has been observed and well studied in the stationary shear flow situation\cite{butler2002factors}. We also observed a shear melting phenomenon in the oscillatory shear deformation, as seen in 
figure \ref{fig:4} (f), which shows that the liquidus temperature $T_L$ decreases as $\gamma_A$ increases. The solid line shows a relation  $\Delta T_L \sim -\gamma_A^2$, and indicates that the shear melting is induced by an additional elastic work associated with the  external loading.
However, the shear melting effect in the case of oscillatory shear is distinct from the case of stationary shear flow.

For stationary shear flow, the strain rate $\dot{\gamma}$ is the only external loading parameter, in contrast, there are two parameters in the oscillatory shear: the  amplitude $\gamma_A$ and the period $t_p$. 
For stationary shear flow, both the thermodynamic and kinetic factors can be modified by a function of strain rate \cite{PhysRevE.87.062307,PhysRevE.99.062801,PhysRevE.93.042803}. 
However,  for  oscillatory shear, the simple combination of these two parameters  can not be simply combined into a strain  rate $\dot{\gamma} \sim \gamma_A/t_p$ does not describe the  changes induced in the nucleation process.
For instance, the TTT curve with $\gamma_A=0.02$ and $t_p=0.02$ ns in Ni$_{50}$Al$_{50}$ is quantitatively similar to the situation with $\gamma_A=0.1$ and $t_p=100$ ns (see Figure \ref{fig:3}(a), (c)).
However, the shear rate of the previous situation is $10^3$ times larger than the latter one.
This means that the shear rate is not an appropriate quantity for the nucleation process under oscillatory shear.

For oscillatory shear, we argue that the elastic stress, rather than shear rate, is the key parameter in crystal nucleation and shear melting.
As is well known, the viscoelastic response of glasses is strongly frequency dependent.
In the frequency range covered by our simulations, the storage modulus for our systems behaves typically 
as an inverse power of frequency, $G'\propto 1/t_p^\alpha$,
with an exponent $\alpha \simeq 0.2$, which is far less than one (Figure \ref{fig:5}), hence the elastic stress can be written  as $\sim \gamma_A/t_p^{\alpha}$. 
As a result, the shear melting can be seen as a stress modified result, to second order in the maximum elastic stress ($\Delta T_L \sim -\gamma_A^2/t_p^{2\alpha}$). 

Following this idea, it is tempting to  introduce an effective temperature $T_{e}(\gamma_A,t_p) \equiv T+T_B \gamma_A^2 (\tau_N^Q/t_{p})^{2\alpha}$ to represent the external perturbation on the shear melting, where $T_B$ is a material parameter. 
We also note that the GFA of alloys measured by $\tau_X$ at the nose is insensitive to the perturbation at first order (figure \ref{fig:4}(b)).
It implies that the derivative of $\tau_X$ with respect to the external perturbation vanishes at the nose.  This observation can be rationalized by assuming that, in a first approximation, the perturbation is described by the effective temperature $T_{e}(\gamma_A,t_p)$ and that the nucleation time is given by $\tau_X(\gamma_A,t_p) = \tau_{X,Q}(T_{e}(\gamma_A,t_p))$ where $\tau_{X,Q}(T)$ is obtained under quiescent conditions. With this assumption, taking the derivative w.r.t. the perturbation (e.g. $\gamma_A$) involves, according to the chain rule,  a derivative w.r.t. temperature, which by construction vanishes at $T_N$. As a result, to first order, the nose temperature will be shifted, but the value of the nucleation time will be stationary. This effective temperature model can be written as:
\begin{equation}
  \ln(\tau_X(\gamma_A,t_p)/\tau_N^Q) ={\hat{W}}(\frac{1}{\hat{T}_e}-1)+ \hat{\Gamma}(\frac{1}{\hat{T}_e (1-\hat{T}_e/\hat{T}_L)^2}-\frac{1}{(1-1/\hat{T}_L)^2})
  \label{eqn:A3}
\end{equation}

Where $\hat{T}_e \equiv T_e/T_N^Q$ is an adimensional effective temperature, and  $\hat{T}_L \equiv {T_{L}}/{T_N^Q}$, $\hat{\Gamma} \equiv {\Gamma}/{k_B T_N^Q}$, $\hat{W} \equiv {W}/{k_B T_N^Q}$ are the fitting parameters of quiescent states by equation \ref{eqn:cnt}.
For Ni$_{50}$Al$_{50}$, $\hat{T}_L= 1.43$, $\hat{W}=19.06$, $\hat{\Gamma}=0.43$ and for Cu$_{23}$Zr$_{77}$ system, $\hat{T}_L=1.31$, $\hat{W}=41.66$, $\hat{\Gamma}=0.40$. 
The parameter $T_B/T_N^Q=2.3 \pm 0.3$ for Ni$_{50}$Al$_{50}$ system is fitted by the simulation result of $\gamma_A=0.1$ and $t_p=0.02$ ns, and $T_B/T_N^Q=3.5 \pm 0.4$ for Cu$_{23}$Zr$_{77}$ system is fitted by the data of $\gamma_A=0.06$ and $t_p=0.02$ ns. 
Interestingly, we take the material parameter $T_B$ into liquidus temperature  $T_L(\gamma_A,t_p)=T_L^Q-T_B \gamma_A^2 (\tau_N^Q/t_p)^{2\alpha}$ , where $T_L^Q$ is the liquidus temperature at the quiescent state, the predicted curve matches pretty well with simulation results (see Figure \ref{fig:4} (f)), it proves that the shear melting can be well depicted by the effective temperature model.
Figure \ref{fig:6} shows the comparison between the TTT curve predicted by equation \ref{eqn:A3} and that of  the simulation data. The effective temperature model can  depict the influence of amplitude and period on the TTT curve in the vicinity of the nose temperature, but it overestimates $\tau_X$ at low temperature  (Figure \ref{fig:6}).
This  implies that the effective temperature model does not represent the kinetic factor well, as the latter  shows a linear dependence on the strain amplitude (Figure \ref{fig:4} (c)).
}
\subsection{\BSS{Phenomenological nucleation model for oscillatory shear}}
\begin{figure}
  \centering
  \includegraphics[width=0.8\textwidth]{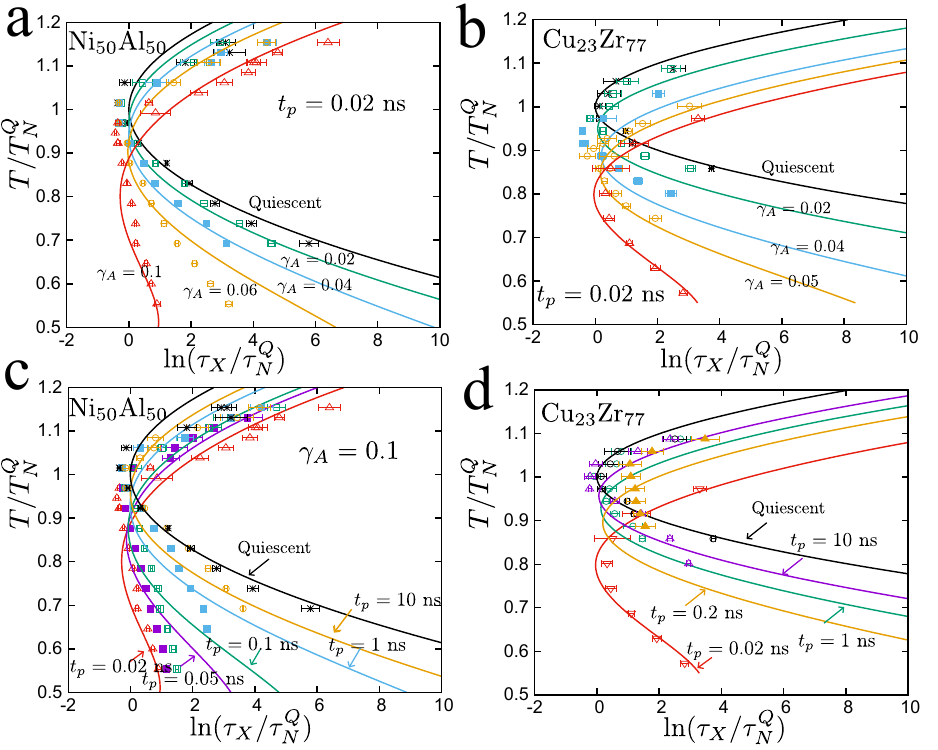}
  \caption{\textbf{Phenomenological model of the TTT curve.}
  \textbf{a,b} The TTT curves change with strain amplitude  $\gamma_A$ for Ni$_{50}$Al$_{50}$ and Cu$_{23}$Zr$_{77}$, respectively, the period is {$t_p=0.02$ ns}.
  \textbf{c,d} The TTT curves change with various periods for Ni$_{50}$Al$_{50}$ with $\gamma_A=0.1$, and for Cu$_{23}$Zr$_{77}$ with $\gamma_A=0.06$. 
  The points  represent the simulation results, and the solid curves are calculated from equation \ref{eq:fit}.
  The three material parameters of the model are fitted from the simulation results 
  of quiescent state for both alloys and $\gamma_A=0.1$ for Ni$_{50}$Al$_{50}$ or $\gamma_A=0.06$ for Cu$_{23}$Zr$_{77}$ with {$t_p=0.02$ ns}.
  }
  \label{fig:7}
\end{figure}
\begin{figure}
  \centering
  \includegraphics[width=0.6\textwidth]{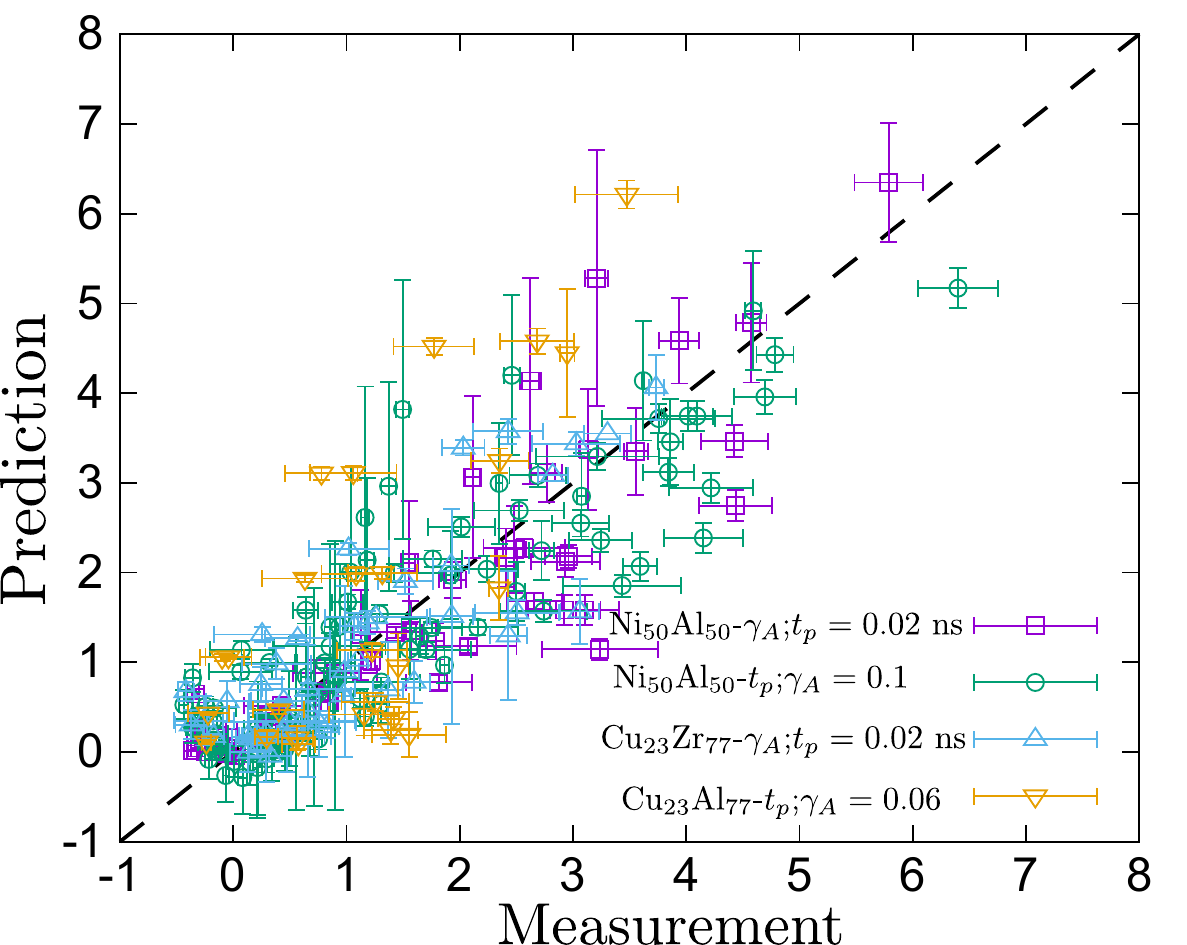}
  \caption{
  The correlation between the phenomenological model prediction and the experiment results of $\ln(\tau_X/\tau_N^Q)$, the dashed line is a guide for the eye.}
  \label{fig:8}
\end{figure}
Based on the above numerical observations, we provide a simple fitting formula  to describe the effect of oscillatory loading on the TTT curve.
We focus on the TTT curve near the nose temperature $T_{N}^Q$ of the quiescent state and the following parameters $T_{N}^Q=1080$ K, $\tau_N^Q=0.579$ ns for Ni$_{50}$Al$_{50}$ and $T_{N}^Q=870$ K, {$\tau_N^Q=12.657$ ns} for Cu$_{23}$Zr$_{77}$.
In the vicinity of the nose, 
the TTT curve of the quiescent state takes the form: 
\begin{equation}
  \ln ({\tau_X}/{\tau^Q_N}) = \hat{A} (T/T_N^Q-1)^2
  \label{eqn:2}
\end{equation}
Where $\tau^Q_N$ is the incubation time in the quiescent state at the nose temperature. $\hat{A}$ is a material parameter 
which can be well fitted using the numerical results in the quiescent state (Figure \ref{fig:1} d), $\hat{A}= 67 \pm 7$ for Ni$_{50}$Al$_{50}$ and $\hat{A}=203 \pm 19$ for Cu$_{23}$Zr$_{77}$. 

When the oscillatory shear is applied during the nucleation, the TTT curve will be modified. \BSS{As discussed above, we found that the relevant parameter is $\gamma_A (\tau_N^Q/t_p)^\alpha$, with $\alpha=0.2$.  Generally, the incubation time is controlled by three factors: the prefactor, kinetic and thermodynamic factors. 
The shortest incubation time results from  the balance between these three factors.
Our analysis shows that the kinetic factor, $W(\gamma_A,t_p)$, decreases linearly with  the strain amplitude (Figure \ref{fig:4} (c)). $W(\gamma_A,t_p) \sim -\gamma_A (\tau_N^Q/t_p)^\alpha$ . Furthermore, there is a good correlation between the prefactor and the kinetic factor at nose temperature ( see Figure \ref{fig:4} (d)) , implying that the change in the prefactor  is also proportional to $\gamma_A (\tau_N^Q/t_p)^\alpha$.
The  thermodynamic factor, it controlled by the liquid-crystal interface parameter $\Gamma$ and liquidus temperature $T_L$ (see equation \ref{eqn:cnt}). Both $\Gamma$ and $T_L$ display a  quadratic dependence  on the strain amplitude  (Figure \ref{fig:4}),  so that at first order, We can take the change thermodynamic barrier as $\Delta G(T) \sim \gamma_A^2 (\tau_N^Q/t_p)^{2\alpha}$. 
Therefore, the incubation time under oscillatory shear can be written as:
}
\begin{equation}
  \label{eq:fit}
  \begin{split}
	\ln(\tau_X/\tau_N^Q)=\hat{A}(T/T_N^Q-1)^2-\hat{B}\gamma_A {\left(\frac{\tau_N^Q}{t_p}\right)^\alpha}(\frac{T_N^Q}{T}-1)+{\hat{C}} \gamma_A^2 \left(\frac{\tau_N^Q}{t_p}\right)^{2\alpha} \frac{T_N^Q}{T}
  \end{split}
\end{equation}
Where $\hat{B}$, $\hat{C}$ are material parameters. We fitted $\hat{B}$ and $\hat{C}$ from the numerical results of $\gamma_A=0.1$, {$t_p=0.02$ ns} for Ni$_{50}$Al$_{50}$ ($\hat{B}=94 \pm 3$, $\hat{C}=34 \pm 4$) and $\gamma_A=0.06$, {$t_p=0.02$ ns} for Cu$_{23}$Zr$_{77}$ ($\hat{B}=239 \pm 2$, $\hat{C}=102 \pm 3$).
Interestingly, this simple formula agrees \emph{quantitatively} well with the simulation data of various strain amplitudes $\gamma_A$ and periods $t_p$ (Figure \ref{fig:7}).  The correlation between the r.h.s and l.h.s of equation \ref{eq:fit} is shown in Figure \ref{fig:8}.

An interesting observation of equation \ref{eq:fit} is that the oscillatory deformation induces a linear decrease of the kinetic factor, and increases the thermodynamic factor quadratically as a function of the stress. \BSS{For $\gamma_A (\tau_N^Q/t_p)^\alpha < 1$, the decrease of the kinetic factor is faster than the increase of the thermodynamic factor, and the nucleation will be  accelerated by the atomic mobility, in contrast,  for $\gamma_A (\tau_N^Q/t_p)^\alpha > 1$, the thermodynamic factor increases more rapidly than the kinetic factor decreases}. Therefore one can on the one hand improve the atomic mobility of the alloy, and on the other hand, the nucleation can be significantly retarded. 
This is important and attractive for the fabrication of metallic glasses and control the nucleation process of the alloy,  one can use ultrasonic vibration to enhance mobility meanwhile avoiding crystallization during forging or cold joining of metallic glasses\cite{Li2014,Ma2019}.

\section{Conclusion}
In conclusion, we find the TTT curve under oscillatory shear can be adjusted  by tuning the amplitude of the strain and its frequency. We provide an empirical  model to describe the  dependence of the TTT curve on the loading, which can quantitatively well represent the numerical data. This formula is interpreted by introducing the elastic stress associated with the loading, which appears to be the essential parameter that will influence the nucleation time.
Our study shows that,  using oscillatory loading, the mobility and nucleation of alloys can be decoupled in the deeply supercooled liquid state, therefore the formation ability can be enhanced and nucleation can be retarded. This is particularly attractive and useful for the manufacturing and processing of metallic glasses.

\section{Acknowledgements}

This work is supported by Guangdong Major Project of Basic and Applied Basic Research, China (Grant No.2019B030302010), 
Guangdong Basic and Applied Basic Research, China (Grant No.2021B1515140005, 2022A1515010347), the NSF of China (Grant Nos.52130108, Nos.U1930402),  B.S.S acknowledges the computational support from the Platform for Data-Driven Computational Materials Discovery of Songshan Lake Materials Laboratory. 

\bibliographystyle{unsrt}
\bibliography{ref.bib}
\section*{Appendix}
\renewcommand{\thefigure}{A\arabic{figure}}
\renewcommand{\theequation}{A\arabic{equation}}
\renewcommand{\thesection}{A\arabic{section}}
\setcounter{figure}{0}
\setcounter{equation}{0}
\setcounter{section}{0}
\section{Effect of simulation parameters on the nucleation process}
\begin{figure}[!htpb]
\centering
\includegraphics[width=0.8\textwidth]{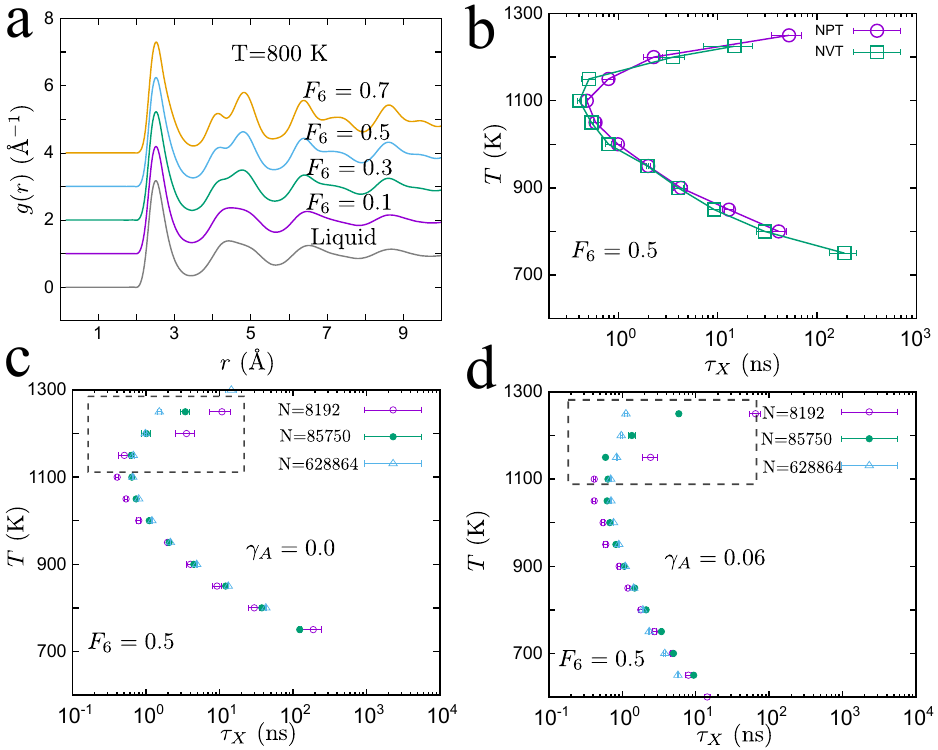}
\caption{
\textbf{{Comparison of }simulation parameters for the TTT curves {for Ni$_{50}$Al$_{50}$}} 
\textbf{a.} The pair distribution function $g(r)$ varies with 
the degree of crystallization $F_6$ at $T=800$ K. 
\textbf{b.} Comparing the NPT and NVT ensemble on the TTT curve for $\gamma_A=0.0$.
\textbf{c.,d.}  The TTT curve of various system sizes ($N=8192$, $85750$, $628864$) with $\gamma_A=0.0$ and $\gamma_A=0.06$, respectively.
}
\label{fig:A1}
\end{figure}
We investigated the role of control parameters such as the NPT ensemble, and system size during the nucleation process.
Figure \ref{fig:A1}(a) shows the pair distribution function $g(r)$ varies with $F_6$ at $T=800$ K, as the degree of crystallization increases, the {alloy} is transforming from liquid state to crystal state.
For the sample quenching from high temperature with the NPT ensemble, the TTT curves with the NPT or the NVT ensemble are equivalent during the nucleation process (Figure \ref{fig:A1} (b)).
Figures \ref{fig:A1} (c) and (d) show that the TTT curves with loading and quiescent condition reveal similar finite-size effects.  
Above $T_N$, the incubation time is notably reduced as the system size increases (the dashed rectangular region), in contrast, below $T_N$, $\tau_X$ is intensive with the system size.
The nucleus formation is spatially localized, and the nucleation rate is intensive to system size, however, as the system size increases, the probability that the {embryo} reaches the critical size is increasing, and above $T_N$, the nucleus will grow rapidly, hence the incubation time will dependent on the system size. 
In contrast, below $T_N$, the nucleus {grows} sluggishly, and the incubation time is insensitive to {the }system size. 

\section{Effect of threshold for the incubation time.}
\begin{figure}[!htpb]
\centering
\includegraphics[width=0.8\textwidth]{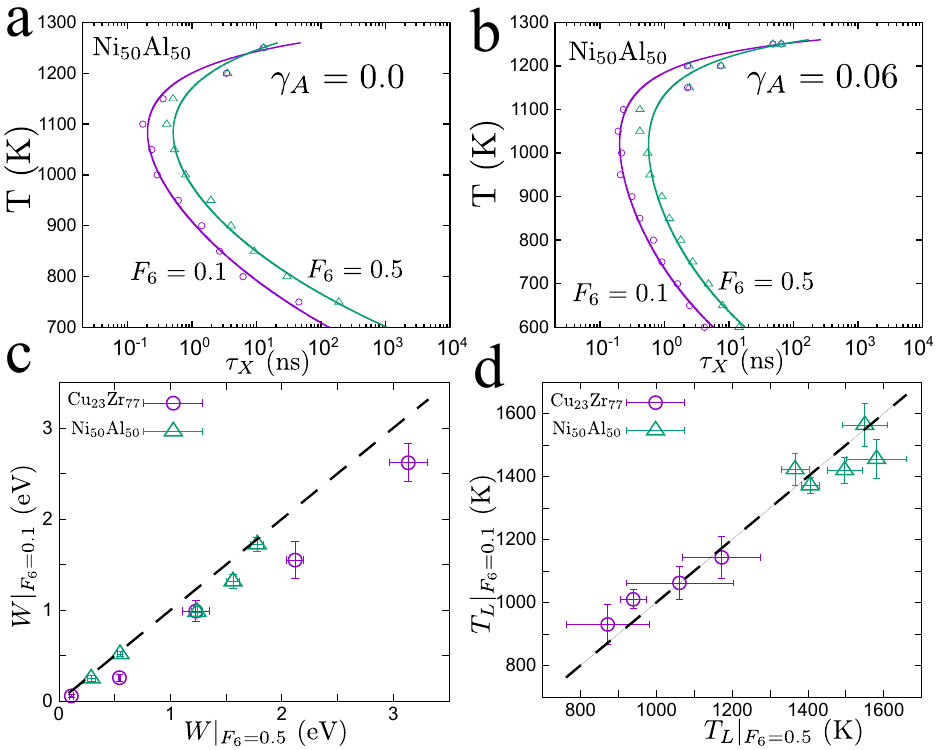}
\caption{
\textbf{{Incubation} time $\tau_X$ versus the degree of crystallization $F_6$}. 
\textbf{a.,b.} Comparing the TTT curve for $F_6=0.1$ or 0.5 of Ni$_{50}$Al$_{50}$, with $\gamma_A=0.0$ and $\gamma_A=0.06$, respectively, {$t_p=0.02$ ns}. 
The solid lines are fitted from the CNT formula with simulation data.
\textbf{c.} The correlation between kinetic barriers $W$ extracted from TTT curve of $F_6=0.1$ or 0.5 with various amplitude strains and {$t_p=0.02$ ns}.
\textbf{d.} The correlation between {liquidus} temperature $T_L$  from TTT curve of $F_6=0.1$ or 0.5 with various amplitude strains and {$t_p=0.02$ ns}.
The dashed lines are guided for the eyes.
}
\label{fig:A2}
\end{figure}
We investigated the sensitivity of the TTT curves on the threshold of the incubation time $\tau_X$.
Figure \ref{fig:A2} (a) and (b) show the TTT curve evolves with the degree of crystallization $F_6$ at $\gamma_A=0.0$ and $\gamma_A=0.06$, respectively. 
The nose temperature $T_N$ is insensitive to the threshold, in contrast, the incubation time $\tau_N$ at $T_N$ is retarded as the degree of crystallization increases.
Above $T_N$, the TTT curve is insensitive to $F_6$ and below $T_N$ the TTT curve will be retarded with increasing $F_6$, 
it indicates the threshold of incubation time mainly impacts the kinetic term of the nucleation process, with more minor influences on the thermodynamic term. 
Figure \ref{fig:A2} (c) shows the correlation between the kinetic barrier of various thresholds, there is slightly deviated from the equivalent line, and the kinetic barrier $W$ with $F_6=0.5$ is larger than $F_6=0.1$, this is consistent with the retardation of nucleation below $T_N$.
Figure \ref{fig:A2} (d) shows the correlation between the temperature $T_L$ of various thresholds, $T_L$ is not changed with threshold and it supports that the thermodynamic term is insensitive to the threshold of incubation time.
\
\end{document}